# Early Development of UVM based Verification Environment of Image Signal Processing Designs using TLM Reference Model of RTL

Abhishek Jain
[1]Imaging Group and [2]Jaypee Business School
[1]STMicroelectronics and [2]Jaypee Institute of Information Technology Greater Noida, India

Dr. Hima Gupta
Jaypee Business School
Jaypee Institute of Information Technology
Noida, India

Sandeep Jana
SDS Group
STMicroelectronics Greater Noida, India

Krishna Kumar
SDS Group
STMicroelectronics Greater Noida, India

*Abstract*—With semiconductor industry trend of "smaller the better", from an idea to a final product, more innovation on product portfolio and yet remaining competitive and profitable are few criteria which are culminating into pressure and need for more and more innovation for CAD flow, process management and project execution cycle. Project schedules are very tight and to achieve first silicon success is key for projects. This necessitates quicker verification with better coverage matrix. Quicker Verification requires early development of the verification environment with wider test vectors without waiting for RTL to be available.

In this paper, we are presenting a novel approach of early development of reusable multi-language verification flow, by addressing four major activities of verification –

1.  **Early creation of Executable Specification**

2.  **Early creation of Verification Environment**

3.  **Early development of test vectors and**

4.  **Better and increased Re-use of blocks**

Although this paper focuses on early development of UVM based Verification Environment of Image Signal Processing designs using TLM Reference Model of RTL, same concept can be extended for non-image signal processing designs.

## I. INTRODUCTION

Image signal processors (ISP) address different markets, including high-end smartphones, security/surveillance, gaming, automotive and medical applications. The use of industry standard interfaces and rich set of APIs makes the integration

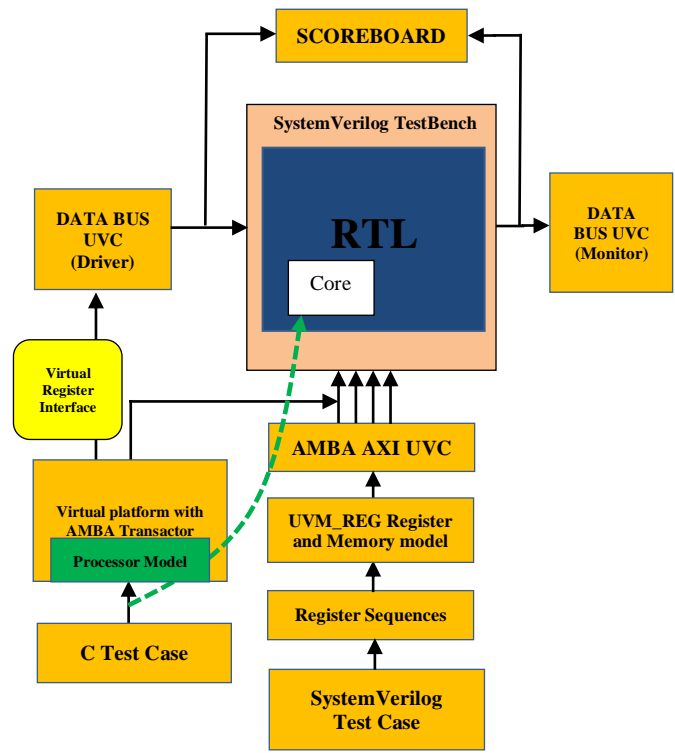

Fig. 1. Verification Environment of Image Signal Processing Design

of image processors a straightforward process and helps to reduce end-product time to market.

Image signal processing algorithms are developed and evaluated using C/Python models before RTL implementation. Once the algorithm is finalized, C/Python models are used as a golden reference model for the IP development. To maximize re-use of design effort, the common bus protocols are defined for internal register and data transfers.

A combination of such configurable image signal processing IP modules are integrated together to satisfy a wide range of complex image signal processing SoCs [1].

In Verification Environment of Image Signal Processing design as shown in figure 1, Host interface path is used to do programming of configurable blocks using SystemVerilog UVM based test cases. UVM_REG register and memory model [20] is used to model registers and memories of DUT. DUT registers are written/read via control bus (AXI3 Bus here) UVC. RTL control bus interface acts as target and control bus UVC acts as initiator. The target control interface of the ISP RTL is driven by control bus UVC (configured as initiator).After register programming is done, image data(random/user-defined) is driven to the data bus interface by the data bus UVC and the same data is also driven to the reference model. Output of the ISP RTL is received by the receiver/monitor of the data bus UVC. Scoreboard compares the output of RTL and reference model and gives the status saying whether the both output matches or not.

'C' test cases are used for programming of RTL registers/memories via CPU interface. C test cases control the SystemVerilog Data Bus UVC using Virtual Register Interface (VRI) [15], [18]. VRI layer is a virtual layer over verification components to make it controllable from embedded software. It gives flexibility to Verification Environment users to use the Verification IPs without knowing SystemVerilog.

Generally, development of Verification Environment for verification of designs is started after availability of the RTL. Thus, significant time is spent for setup and debugging of verification environment after release of RTL which results in delay in start and completion of verification of the designs. It is required to find ways to start developing the Verification Environment much before the arrival of the RTL so that when RTL is available, Verification Environment can be easily plug and play and verification of the designs can be started quickly. Use of TLM reference model of RTL for development of Verification Environment much before arrival of RTL proves to be good solution for the above mentioned problem.

This paper is focusing on early development of UVM based Verification Environment of Image Signal Processing designs using TLM Reference Model of RTL before availability of the RTL. Early development of Verification Environment of Image Signal Processing designs is described in detail in Section II.

## II. EARLY DEVELOPMENT OF UVM BASED VERIFICATION ENVIRONMENT

### A. Modeling of ISP designs

A loosely timed high level model of the ISP block is generated at algorithmic functional level using C/C++/SystemC and with TLM-2 interface.

The SCML – SystemC Modeling Library, an open source SystemC library from Synopsys Inc. [26] is being used here.

The purpose of this model generation is to use this as a reference model. We may say it as a "Golden Reference Model" or "Executable Functional Specification" of the ISP designs. From functional and structural perspective this model can be divided in two major spaces.

**First space -** the algorithmic computational part, is mainly responsible for image processing using various algorithms involved for image manipulation from the incoming image stream data.

**The second space** – a TLM interface, is responsible for all kinds of communication to external IPs and other system blocks.

Register interface of this model is generated using IP-XACT tools. And algorithmic part is manually implemented.

### B. Testing of Executable Spec only

To test the TLM ISP model, an environment is developed using Python (an open source scripting language) and Synopsys Pa-Virtualizer Tool Chain.

The test environment has following major components:

- Test bench in Python
- Configuration file reader in Python
- Raw Data Reader
- ISP model
- Input data injector in Python
- Output data receiver in Python
- Output data checker in Python
- Synopsys Pa-Virtualizer Tools Chain for GUI, debugging, and simulation

XML file format is used for test bench configuration and passing other parameter to the testing environment.

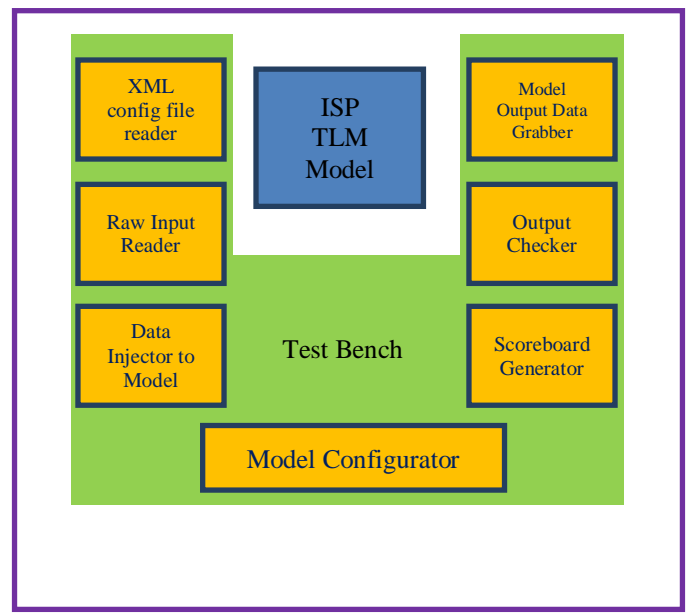

Fig. 2. ISP Model Testing Environment

## C. Use of TLM ISP Model for early development of RTL Verification Environment

After the ISP model is proved to be functionally correct, the same model is used for early development of RTL functional verification environment.

A suitable TLM sub-system is designed. This TLM sub-system consists of various models namely; ISP functional model, AXI BFM, configurable clock generators model, configurable reset generator model, memory model, configurable interconnect etc. All these are pure SystemC models. AXI BFM is provided to interact with other part of the world.

ISP RTL block needs exhaustive verification, which is possible only when the RTL is ready. But, development of RTL design takes time, which means verification of RTL design can't be possible before it becomes available. To shorten this sequential activity, functional model of ISP is used to prepare the early verification environment.

A SystemVerilog test bench wrapper is created over SystemC/TLM ISP sub-system. This SystemVerilog test bench interface with the RTL verification environment.

## D. Virtual Platform Sub-system

When all components of platform are in TLM/C, means C/C++ are used as modeling language; we call it a Pure Virtual platform. In typical verification environment, generally all verification components are not only TLM based but also of different verification languages thus making it a Multi-language heterogeneous simulation environment. For developing early verification environment, TLM based sub-system is developed which consists of every block in TLM/C. This TLM based Sub-system is model of RTL.

In the above mentioned RTL verification environment, a processor model is used which enables us to early develop 'C' test cases for programming of RTL registers/memories via CPU interface. The challenge is to keep the verification environment independent of "C" test cases. We don't wish to compile every time whenever there is change in application code. To be able to achieve this, a sub-system is designed which consists of models of bus interfaces, like AXI BFM, a "generic" processor model, model of memory, etc. an independent "C" program/test case is written to do all the programming and configuration, which in turn runs on processor model of this sub-system. This sub-system is active element in programming phase, but becomes passive once the programming is complete.

Virtual platform sub-system can be represented in following block diagram.

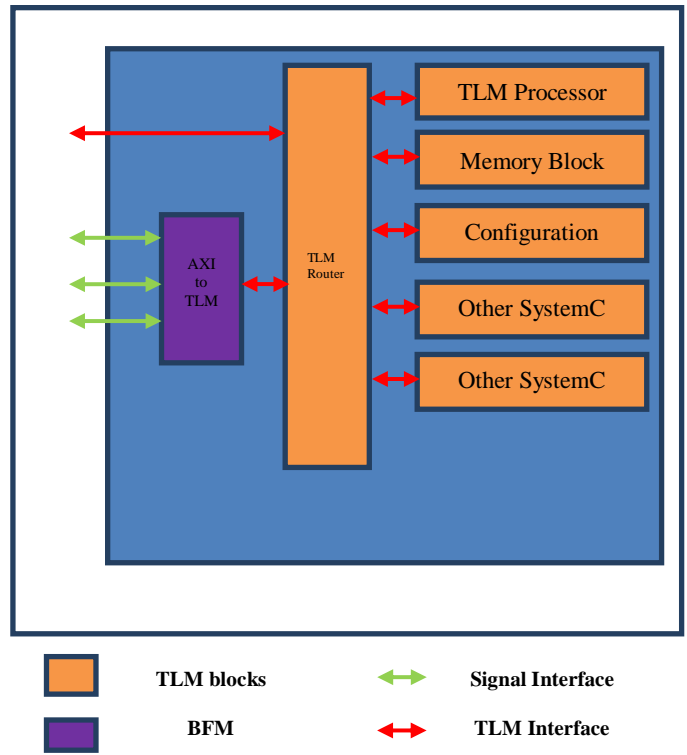

Fig. 3. Virtual Platform Sub-system

## E. Virtual Register Interface (VRI)

Today, most of the embedded test infrastructure uses some adhoc mechanism like "shared memory" or synchronization mechanism for controlling simple Bus functional models (BFMs) from embedded software.

In order to provide full controllability to the "C" test developer over these verification components, a virtual register interface layer is created over these verification environments which provides the access to the sequences of these verification environment to the embedded software enabling configuration and control of these verification environments to provide the same exhaustive verification at SoC Level.

This approach addresses the following aspects of verification at SoC Level:

➢ Configuration and control of verification components from embedded software.
➢ Reusability of verification environments from IP to SoC.
➢ Enables reusability of testcases from IP to SoC.
➢ Providing integration testcases to SoC team which is developed by IP verification teams.

It has been achieved by using Virtual Register Interface (VRI) layer over Verification components [18]. VRI layer over verification components is –

➢ A virtual layer over verification environment to make it controllable from embedded software

➢ Provides high level C APIs hiding low level implementation

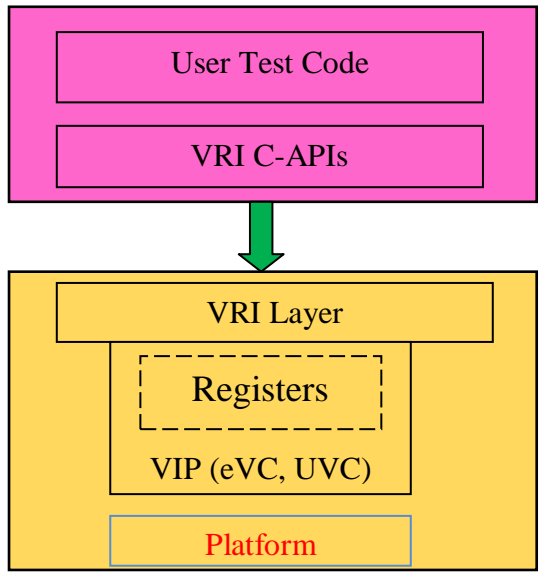

Fig. 4.  Virtual Register Interface (VRI)

An example of C test case using VRI interface is as follows –

*vr_enet_packet pkt;*
*vr_enet_packet rx_pkt;*
*rx_pkt.data = new vri_uint8_t[2000]; //create buffer for receiving data*

*pkt.packet_kind = ETHERNET_802_3;*
*pkt.data_length = 0; //RANDOM DATA*
*pkt.dest_addr_high = 0x11ff;*
*pkt.src_addr_high = 0x2288;*
*pkt.tag_kind = UNTAGGED;*
*pkt.tag_prefix = 0x1234;*
*pkt.s_vlan_tag_prefix = 0x5678;*
*pkt.err_code = 0;*
*for (int i=0;i<100;i++) {*
 *pkt.dest_addr_low = i;*
 *pkt.src_addr_low = i+1;*
 *enet_send_pkt(0,&pkt);       //send packet to ENET UVC instance0 (MAC)*
 *enet_recv_pkt(1,&rx_pkt); //receive packet from ENET UVC instance1 (PHY)*
 *compare_pkt(pkt,rx_pkt);*
 *};*

*F.  Flow used for Design Verification*

Much before arrival of RTL, C/Python model of image signal processor designs is developed for algorithm evaluation. Then, TLM/SystemC model of the design is created from C/Python model. After proper exhaustive validation of the model with required test vectors, the model qualifies as an Executable Golden Model or Executable Specification means a 'living' benchmark for design specification. Enabling the use of TLM Model as DUT expedites development and better proofing of the verification environment with wider test vectors without waiting for RTL to be available.

Standard 'interfaces' are used to enable the reuse of verification components. In addition to standard method of bus-interface or signals level connectivity, UVM Multi-Language Open Architecture is used to connect System Verilog TLM port directly to SystemC TLM port which gives advantage of better simulation speed and better development/debug cycle in addition of clean, better and easy connectivity/integration of blocks. Presence of TLM components gives us flexibility to make backdoor direct access to the DUT registers and memories.

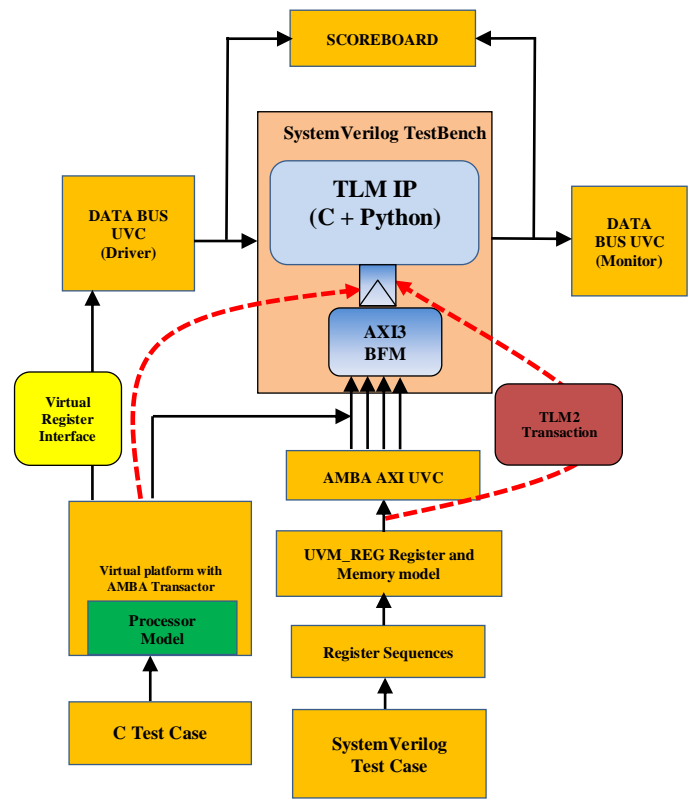

Fig. 5.  Early development of Verification Environment using TLM Model

A processor model is used which enables us to early develop 'C' test cases for programming of RTL registers/memories via CPU interface. Same 'C' test cases are used for controlling the SystemVerilog UVC's using Virtual Register interface (VRI) layer. In our verification environment, alternative Host interface path is used to do programming of configurable blocks using SystemVerilog UVM based test cases.

In both above cases, control/data flows across both TLM and bus interface boundaries. This method enhances the chances of re-using different already existing blocks in flow. IP-XACT based tools are also used for automatically configuring the environment for various designs.

By the time RTL arrives, complete verification environment and test-vectors are ready with sufficient sanctity, thus eliminating the number of verification environment issues which may arise when actual RTL verification is started. When RTL arrives, the TLM/SystemC model is simply replaced with RTL block with reuse of maximum of other verification components. This enhances the rapid/regress testing of design immediately. Also same C test cases can be run on actual core.

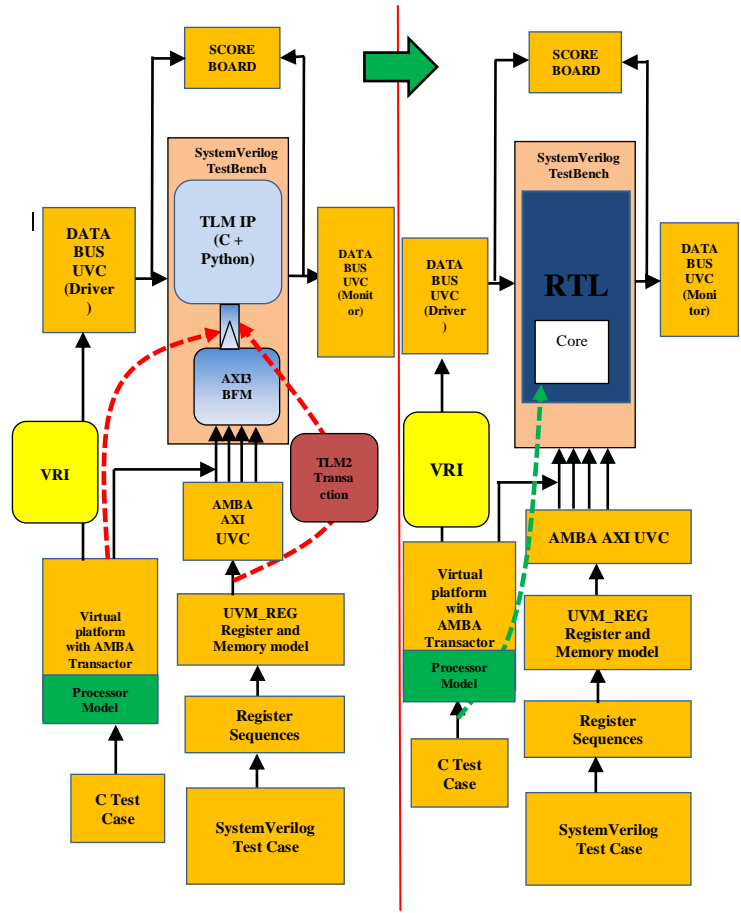

Fig. 6.   : Reuse of early developed Verification Environment

## III.  RESULTS

Using TLM reference model of the RTL, UVM based Environment for verification of design is developed without waiting for RTL to be available. Significant reduction in overall verification time of the design is achieved.

## IV.  CONCLUSIONS

TLM/SystemC reference model of the design is the key component to enable the early development of Verification Environment without waiting for RTL to be available. UVM based early verification Environment is developed using TLM/SystemC reference model of the design. Verification

Environment is developed both with Host interface and Core using Virtual Register Interface (VRI) approach. IP-XACT based tools are used for automatically configuring the Verification Environment. Testing of features of Verification Environment at TLM abstraction level runs faster and thus, it overall speeds up functional verification. Same environment can be reused from IP level to SOC level or from one SOC to another SOC with no/minimal change. Verification Environment is reusable both vertically and across projects thus saving further time across projects.

### ACKNOWLEDGMENT

The authors would like to specially thank to their management Giuseppe Bonanno (CAD Manager, Imaging Division, and STMicroelectronics) and Antoine Perrin (Manager, SDS Team, STMicroelectronics) for their guidance and support. We would also like to thank management and team members of Imaging Division, STMicroelectronics; Faculty members and peer scholars of JBS, Jaypee Institute of Information Technology University and also Cadence team for their support and guidance.

## AUTHOR'S PROFILE

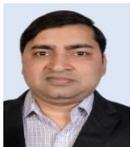

**Abhishek Jain, Technical Manager, STMicro-electronics Pvt. Ltd.**
**Research Scholar, JBS, Jaypee Institute of Information Technology, Noida, India.**
**Email: ajain_design@yahoo.co.in;**
**abhishek-mmc.jain@st.com**

Abhishek Jain has more than 11 years of experience in Industry. He is driving key activities on Functional Verification Flow in Imaging Division of STMicroelectronics. He has done PGDBA in Operations Management from Symbiosis, M.Tech in Computer Science from IETE and M.Sc. (Electronics) from University of Delhi. His main area of Interest is Project Management, Advanced Functional Verification Technologies and System Design and Verification especially UVM based Verification, Emulation/Acceleration and Virtual System Platform. Currently, he is doing Research in Advanced Verification Methods for Efficient Verification Management in Semiconductor Sector. Abhishek Jain is a member of IETE (MIETE).

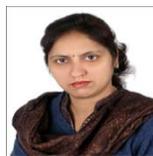

**Dr. Hima Gupta, Associate Professor, Jaypee Business School (A constituent of Jaypee Institute of Information Technology University), A – 10, Sector-62, Noida, 201 307 India.**
**Email: hima_gupta2001@yahoo.com**

Dr. Hima has worked with LNJ Bhilwara Group & Bakshi Group of Companies for 5 yrs. and has been teaching for last 11 years as Faculty in reputed Business Schools. She also worked as Project Officer with NITRA and ATIRA at Ahmedabad for 5 years.

She has published several research papers in National & International journals.

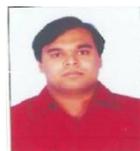

**Sandeep Jana, Staff Engineer, STMicroelectronics Pvt. Ltd.**
**Email: sandeep.jana@st.com**

Sandeep Jana is Staff Engineer at STMicroelectronics managing the TLM based Verification activities at Greater Noida. He has an expertise of over seven years in various aspects of ESL domain such as TLM modeling, Architectural exploration, Platform Integration, Mixed language Platforms, Advanced Verification Methodologies etc. He has been with ST since last 6 years and was previously working in VLSI group of HCL Technologies in their ESL domain. He has a B.Tech degree in Electronics Engineering from MDU Rohtak.

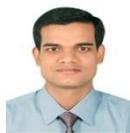

**Krishna Kumar, STMicroelectronics Pvt. Ltd.**
**Email: krishna.kumar@st.com**

Krishna Kumar with almost 12 years at STMicro-electronics has experience in ESL and Placement and Routing of FPGA software tool chain. He holds B.Tech degree in Computer Engineering from Aligarh Muslim University, India.